\documentclass[11pt,letterpaper]{article}
\begin{document}
\title{The extended Lorentz force}
\author{Michael A. Graber
     \\6741 Old Waterloo Road\\Elkridge, Maryland 21075 USA\\
      \footnotesize{e-mail: MikeGraber@SprintMail.com} }
\date{}
\maketitle

\begin{abstract}
The Lorentz force equations provide a partial description of the geodesic
motion of a charged particle on a four-manifold.  Under the hypothesis that 
Maxwell's equations express symmetry properties of the Ricci tensor,
the full electromagnetic connection is determined.  From this connection,  
the fourth equation of the geodesic is derived. 
The validity of this fourth equation can be determined by studying 
the decay of charged particles in an electric field. 
Time will accelerate or decelerate relative to the proper time of a charged 
particle moving in an electric field.  Unstable charged particles 
moving in opposite directions 
parallel to an electric field should exhibit different decay rates.
\end{abstract}

It has been proposed before that the motion of a charged particle 
in an electromagnetic field on the 
four-manifold of our universe is described by the geodesic equation. 
(See Kaluza 1921, Klein 1926, Fock 1927, which are translated and 
reprinted in O'Raifeartaigh 1997.)  At this time, it remains unclear as to
whether such motion does follow a geodesic path.  

The geodesic equation on a four-manifold is four equations giving 
the second derivatives of the four spatial and temporal coordinates:
\[  \frac{d^2\, x}{ds^2 }\: , \qquad \frac{d^2\, y}{ds^2 }\: , \qquad \frac{d^2\, z}{ds^2 }\: , \qquad \frac{d^2\, t}{ds^2 }\: , 
\]
where $s$ is the geodesic length parameter.   
The Lorentz force equations are an empirical solution for 
the motion of a charged particle in three-space; they 
account for only three of the second derivatives.  It is 
possible that $d^{2}t/ds^{2}$ is zero.  
However, there is no theoretical justification for this 
assumption.  The Lorentz force equations were constructed to 
explain experimental results.  At the time the equations were developed, 
there was no motivation to look beyond three dimensions. 

This analysis develops a geometric model based on the hypothesis that 
Maxwell's equations express symmetry properties of the Ricci tensor. 
The full electromagnetic connection is derived.  The connection 
components determine the geodesic equation. 
Defining the proper time $\tau$ in terms of the length parameter, $c\, \tau=s$, 
the model determines the fourth equation of the Lorentz force
to be  
\[  \frac{d^2\, t}{d\tau^2 }= 2 \frac{q}{mc^2} \left( \mathbf{ E\, \cdot} 
    \frac{d\mathbf{x}}{d\tau} \right)  \,\frac{dt}{d\tau}\, 
\]
This equation can be approximated as 
\[  \frac{d^2\, t}{d\tau^2 }= 2 \frac{q}{mc^2} \, \mathbf{ E\, \cdot\, \mathbf{v}} \, 
\]
The new ``force'' equation can be subjected to experimental verification 
by studying the decay 
of charged particles in an electric field. 
Time will accelerate or decelerate relative to the proper time of a charged 
particle moving in an electric field.  Unstable charged particles 
moving in opposite directions 
parallel to an electric field should exhibit different decay rates.

\section{Development of the Connection}

The classical Lorentz force equation 
\[ m\, \mathbf{a} = q \mathbf{E} +\frac{q}{c} \mathbf{v}\times \mathbf{B}  
\]
provides an initial source of information about the connection components.
The form of the Lorentz force appears to contradict 
its interpretation as a geodesic equation, which must be uniformly 
quadratic in the components of the 4-velocity. However, the 
fourth component of the 4-velocity, $d(c\,t)/ds$, would be very close to 1 
and essentially constant in classical experiments; it would not 
have been observable.  In presenting the Lorentz force as 
a first approximation to the geodesic path, we add the fourth velocity component 
to the classical terms 
by writing them as 
\[ \frac{d^{2}x}{ds^{2}} =  \frac{q}{mc^2} E_x \frac{d(c\,t)}{ds} \frac{d(c\,t)}{ds} 
+  \frac{q}{mc^2} B_z \frac{dy}{ds} \frac{d(c\,t)}{ds} 
-  \frac{q}{mc^2} B_y \frac{dz}{ds} \frac{d(c\,t)}{ds} 
\]
\[ \frac{d^{2}y}{ds^{2}} =  \frac{q}{mc^2} E_y \frac{d(c\,t)}{ds} \frac{d(c\,t)}{ds} 
+  \frac{q}{mc^2} B_x \frac{dz}{ds} \frac{d(c\,t)}{ds} 
-  \frac{q}{mc^2} B_z \frac{dx}{ds} \frac{d(c\,t)}{ds} 
\]
\begin{equation} \label{lGeod}
 \frac{d^{2}z}{ds^{2}} =  \frac{q}{mc^2} E_z \frac{d(c\,t)}{ds} \frac{d(c\,t)}{ds} 
+  \frac{q}{mc^2} B_y \frac{dx}{ds} \frac{d(c\,t)}{ds} 
-  \frac{q}{mc^2} B_x \frac{dy}{ds} \frac{d(c\,t)}{ds} 
\end{equation}
This step creates the quadratic structure that allows the interpretation of the Lorentz 
force as a geodesic.

The electromagnetic equations presented here use Gaussian units, 
as described in Jackson (1962) p.~616.  
The geodesic equation in terms of the connection components $\Gamma^{i}{}_{\!jk}$ is 
\[ \frac{d^{2}x_{i}}{ds^{2}} +\sum_{j,k} \Gamma^{i}{}_{\!jk} \frac{dx_{j}}{ds} \frac{dx_{k}}{ds} =0
\]
The connection components appear in equation \ref{lGeod} as the negative 
of the coefficients of the derivative terms on the right-hand side.

There is a certain degree of ambiguity in the initial placement of the $B$-valued components.  
After studying different placements, the components from the classical Lorentz force 
equation were placed in the following manner: 

\begin{center} Table 1\\ Initial Lorentz Components \\[0.5ex]

\begin{tabular}{cccc|cccc|cccc}

 \textit{i} & \textit{j} & \textit{k} & $\Gamma^{\,i}{}_{\!jk}$ & \textit{i} & \textit{j} 
    & \textit{k} & $\Gamma^{\,i}{}_{\!jk}$ 
    & \textit{i} & \textit{j} & \textit{k} & $\Gamma^{\,i}{}_{\!jk}$ \\[0.5ex]
\hline

 & & & &   & & & &   & & & \\
1 & 0 & 0 & $ -qE_x/mc^2 $    &  2 & 0 & 0 &  $  -qE_y/mc^2 $   & 3 & 0 & 0 & $  -qE_z/mc^2$ \\ [0.2ex]
2 & 3 & 0 & $ -qB_x/mc^2 $     & 3 & 1 & 0 & $ -qB_y/mc^2 $  & 1 & 2 & 0 &  $ -qB_z/mc^2$ \\[0.2ex]
3 & 0 & 2 &  $ qB_x/mc^2  $    & 1 & 0 & 3 & $  qB_y/mc^2  $  & 2 & 0 & 1 &  $  qB_z/mc^2$ \\

\end{tabular}

\end{center}
\noindent The time dimension is denoted by 0, while 1, 2, 3 are the spatial dimensions x, y, z.

The following alternative placement has identical properties in all areas considered 
in this paper.  

\begin{center} Alternative Lorentz Components \\[0.5ex]

\begin{tabular}{cccc|cccc|cccc}

 \textit{i} & \textit{j} & \textit{k} & $\Gamma^{\,i}{}_{\!jk}$ & \textit{i} & \textit{j} 
    & \textit{k} & $\Gamma^{\,i}{}_{\!jk}$ 
    & \textit{i} & \textit{j} & \textit{k} & $\Gamma^{\,i}{}_{\!jk}$ \\[0.5ex]
\hline

 & & & &   & & & &   & & & \\
1 & 0 & 0 & $ -qE_x/mc^2 $    &  2 & 0 & 0 &  $  -qE_y/mc^2 $   & 3 & 0 & 0 & $  -qE_z/mc^2$ \\ [0.2ex]
2 & 0 & 3 & $ -qB_x/mc^2 $     & 3 & 0 & 1 & $ -qB_y/mc^2 $  & 1 & 0 & 2 &  $ -qB_z/mc^2$ \\[0.2ex]
3 & 2 & 0 &  $ qB_x/mc^2  $    & 1 & 3 & 0 & $  qB_y/mc^2  $  & 2 & 1 & 0 &  $  qB_z/mc^2$ \\

\end{tabular}

\end{center}
\noindent Other alternative placements of the $B$-valued components 
(including one which is a linear combination of the two examples above) give 
good results at this early stage.  However, they have problems 
generating  the fully developed Maxwell's equations.

The full connection is produced using a geometric model 
based on the hypothesis that 
Maxwell's equations express symmetry properties of the Ricci tensor. 
The Ricci tensor is commonly assumed to be symmetric. 
However, this assumed symmetry depends on the Torsion being zero. 
In the present calculation, the Torsion is not required to be 
zero.  In fact, the Torsion is shown to be related to the 
magnetic field.  Therefore, the Ricci tensor does not have 
a simple symmetry.

The Ricci tensor is developed to have the following symmetry:

\noindent 1) the trace of the Ricci tensor is zero,  $R_{ii}=0$, and 

\noindent 2) the off-diagonal components of the Ricci tensor have the 
mixed symmetry
\[R_{0i} + R_{i0} =0
\]
\[R_{ij} - R_{ji} =0
\]
where $0$ denotes the time dimension and $i,j$ denote spatial dimensions.
This hypothesis is \emph{not} coordinate invariant.  The mixed symmetry simply indicates 
that the universe \emph{does} distinguish between time and space.  This 
symmetry has the flavor of the Minkowski metric.  Note that the Minkowski 
metric has not been used in the calculation of the electromagnetic connection. 
No metric was assumed.

The components of the curvature tensor in terms of the connection 
components are
\begin{equation} \label{defCurv}
R^{\,i}{}_{\!jkl} = \left( \partial\Gamma^{i}{}_{\!lj}/\partial x_{k} 
 - \partial\Gamma^{i}{}_{\!kj}/\partial x_{l} \right) 
+\sum_{m} \left( \Gamma^{m}{}_{\!lj} \,\Gamma^{i}{}_{\!km} -\Gamma^{m}{}_{\!kj} \,\Gamma^{i}{}_{\!lm}  \right).
\end{equation}
The Ricci tensor is defined in terms of the curvature as 
$R_{ij} = \sum_{k} R^{\,k}{}_{\!jki}$.

On evaluating the Ricci tensor expressions with the connection components from the 
Lorentz force equation in Table 1, the results are
\[R_{ii} \qquad \Longrightarrow \qquad  \frac{q}{mc^2} {B^2}  - \nabla \mathbf{ \cdot E} 
\]
\[R_{0i} + R_{i0} \qquad \Longrightarrow \qquad \nabla\times \mathbf{B} 
\]
\[R_{ij} - R_{ji} \qquad \Longrightarrow \qquad \mathbf{0} \, 
\]
where a factor of $q/mc^2$ has been dropped.
This result led to the expectation that the trace $R_{ii}$ would generate Coulomb's law and that 
$R_{0i} + R_{i0}$ would generate Ampere's law.  Note that these expressions have 
been evaluated with only the components that can be read from the Lorentz force 
equations.  

Using these observations as guides, the connection was completed. 
The manner in which Maxwell's equations are satisfied is described 
below.
The natural coordinate system for electrodynamics 
is Cartesian coordinates $(ct,x,y,z)$.   
The full connection 
is shown in Table 2, with the indices $(0,1,2,3)$ representing  
the four coordinates.

\pagebreak[4]
\begin{center}Table 2\\ Components of the Electromagnetic connection $\Gamma$\\[0.5ex]

\begin{tabular}{cccc|cccc|cccc}

 \textit{i} & \textit{j} & \textit{k} & $\Gamma^{\,i}{}_{\!jk}$ & \textit{i} & \textit{j} 
    & \textit{k} & $\Gamma^{\,i}{}_{\!jk}$ 
    & \textit{i} & \textit{j} & \textit{k} & $\Gamma^{\,i}{}_{\!jk}$ \\[0.5ex]
\hline

 & & & &   & & & &   & & & \\
1 & 0 & 0 & $ -qE_x/mc^2 $    &  2 & 0 & 0 &  $  -qE_y/mc^2 $   & 3 & 0 & 0 & $  -qE_z/mc^2$ \\ [0.2ex]
2 & 3 & 0 & $ -qB_x/mc^2 $     & 3 & 1 & 0 & $ -qB_y/mc^2 $  & 1 & 2 & 0 &  $ -qB_z/mc^2$ \\[0.2ex]
3 & 0 & 2 &  $ qB_x/mc^2  $    & 1 & 0 & 3 & $  qB_y/mc^2  $  & 2 & 0 & 1 &  $  qB_z/mc^2$ \\[0.2ex]

0 & 3 & 2 &  $ -\frac{1}{2}\;qB_x/mc^2 $   & 0 & 1 & 3 &  $ -\frac{1}{2}\;qB_y/mc^2 $   & 0 & 2 & 1 &  
   $ -\frac{1}{2}\;qB_z/mc^2$ \\[0.2ex]
0 & 2 & 3 &  $  \frac{1}{2}\;qB_x/mc^2 $   & 0 & 3 & 1 & $   \frac{1}{2}\;qB_y/mc^2 $   & 0 & 1 & 2 &  
    $  \frac{1}{2}\;qB_z/mc^2 $\\[0.2ex]

0 & 0 & 1 &  $ -qE_x/mc^2 $   & 0 & 0 & 2 &  $ -qE_y/mc^2  $ & 0 & 0 & 3 &   $ -qE_z/mc^2 $ \\[0.2ex]
0 & 1 & 0 &  $ -qE_x/mc^2  $  & 0 & 2 & 0 &  $ -qE_y/mc^2 $  & 0 & 3 & 0 &   $ -qE_z/mc^2 $ \\[0.2ex]
2 & 2 & 1 &  $   i\sqrt{5/6}\;\,qE_x/mc^2 $   & 3 & 3 & 2 & $   i\sqrt{5/6}\;\,qE_y/mc^2 $  & 2 & 2 & 3 & 
   $   i\sqrt{5/6}\;\,qE_z/mc^2  $ \\[0.2ex]
2 & 1 & 2 &  $   i\sqrt{5/6}\;\,qE_x/mc^2 $  & 3 & 2 & 3 &   $ i\sqrt{5/6}\;\,qE_y/mc^2  $ & 2 & 3 & 2 &  
   $  i\sqrt{5/6}\;\,qE_z/mc^2  $  \\[0.2ex]
3 & 3 & 1 &  $ - i\sqrt{5/6}\;\,qE_x/mc^2 $  & 1 & 1 & 2 &  $  - i\sqrt{5/6}\;\,qE_y/mc^2 $  & 1 & 1 & 3 &  
   $ - i\sqrt{5/6}\;\,qE_z/mc^2  $  \\[0.2ex]
3 & 1 & 3 &  $ - i\sqrt{5/6}\;\,qE_x/mc^2  $ & 1 & 2 & 1 &  $  - i\sqrt{5/6}\;\,qE_y/mc^2  $ & 1 & 3 & 1 & 
   $  - i\sqrt{5/6}\;\,qE_z/mc^2  $  \\[0.3ex]
1 & 2 & 2 & {\small$   \!\! \left(1+i\sqrt{5/6}\right)  \!qE_x/mc^2 $}  & 2 & 3 & 3 &  
   {\small$  \! \! \left(1+i\sqrt{5/6}\right) \!qE_y/mc^2 $ } & 3 & 2 & 2 &   
   {\small$  \! \!\left(1+i\sqrt{5/6}\right) \!qE_z/mc^2$}   \\[0.4ex]
1 & 3 & 3 &   {\small $  \!\! \left(1-i\sqrt{5/6}\right)  \!qE_x/mc^2  $} & 2 & 1 & 1 &  
   {\small$  \! \! \left(1-i\sqrt{5/6}\right) \!qE_y/mc^2 $ } & 3 & 1 & 1 &   
   {\small$ \! \! \left(1-i\sqrt{5/6}\right) \!qE_z/mc^2  $} \\

\end{tabular}

\end{center}

No metric was used to construct the electromagnetic connection.  
A metric consistent with this connection almost certainly exists.  At this 
time the metric is not known.

\subsection{${\mathbf{R_{ii}}}$: Coulomb's Law}

The trace of the Ricci tensor $R_{ii}$ produces 
\[
{ \left( \frac{q}{mc^2} \right) }^2 
(B_x{}^{\!2} + B_y{}^{\!2} + B_z{}^{\!2} +  E_x{}^{\!2} 
+ E_y{}^{\!2} + E_z{}^{\!2}) 
+ 2  \, \frac{q}{mc^2} \, \left( \frac {\partial }{\partial x} \,E_x 
+        \frac {\partial }{\partial y} \,E_y
 +       \frac {\partial }{\partial z} \, E_z    \right).
\]
Setting $R_{ii}=0$ and using the definition for the charge density $\rho$\,, one finds 
\[   4\pi\rho = \nabla \mathbf{ \cdot E} = -  \frac{q}{mc^2} \,
 \frac{ (E^2 +B^2)}{2 } \]
\[ \rho= -\frac{q\,u}{mc^2} \]
where $u$ is the electromagnetic 
energy density $(E^2 +B^2)/8\pi$.

Note that this equation does \emph{not} say that charge density is 
present whenever there is electromagnetic energy.  In  
relativity theory, rest mass is generated by energy; but there can be 
energy in a form other than rest mass.  In this geometric electromagnetic theory, 
too, there can be 
electromagnetic energy without charge density.  But the geometry 
says that whenever there is charge density there is underlying
electromagnetic energy.

\subsection{${\mathbf{R_{0i} + R_{i0}}}$: Ampere's Law with Displacement Current}

The following three sums of off-diagonal components of  the Ricci tensor 
 ($R_{01} + R_{10}$), ($R_{02} + R_{20}$), and ($R_{03} + R_{30}$)
produce the three expressions 
\[
{ \left( \frac{q}{mc^2} \right) }^2 
(E_y B_z -E_z B_y) 
+  \frac{q}{mc^2} \, \left( \frac {\partial }{\partial y} \,B_z
-        \frac {\partial }{\partial z} \,B_y
 -      \frac{1}{c} \frac {\partial }{\partial t} \, E_x    \right),
\]
\[
{ \left( \frac{q}{mc^2} \right) }^2 
(E_z B_x -E_x B_z) 
+  \frac{q}{mc^2} \, \left( \frac {\partial }{\partial z} \,B_x
-        \frac {\partial }{\partial x} \,B_z
 -      \frac{1}{c} \frac {\partial }{\partial t} \, E_y    \right),
\]
\[
{ \left( \frac{q}{mc^2} \right) }^2 
(E_x B_y -E_y B_x) 
+  \frac{q}{mc^2} \, \left( \frac {\partial }{\partial x} \,B_y
-        \frac {\partial }{\partial y} \,B_x
 -      \frac{1}{c} \frac {\partial }{\partial t} \, E_z    \right).
\]
Setting each of the three expressions to $0$ and using the 
definition for the current density $\mathbf{J}$\,, one finds 
\[   \frac{4\pi}{c} \mathbf{J} = \nabla\times \mathbf{B} -\frac{1}{c} 
  \frac {\partial }{\partial t} \, \mathbf{E}
 = -  \frac{q}{mc^2} \,
  \mathbf{E} \times \mathbf{B} \]
\[ \mathbf{J} = -\frac{q\, \mathbf{S}}{mc^2} \]
where $\mathbf{S}$ is Poynting's vector for the electromagnetic energy flow 
$(c /4 \pi)\,(\mathbf{E} \times \mathbf{B})$.

As in the discussion of Coulomb's law above, this equation does \emph{not} 
assert that whenever there is a propagating electromagnetic wave or a non-zero 
Poynting vector there must 
be a charge density or a current.  Rather, the geometry says that a current is generated 
by an underlying Poynting vector

\subsection{${\mathbf{R_{ij} - R_{ji}}}$: Faraday's Law}

If we evaluate the differences in off-diagonal components 
 ($R_{12} - R_{21}$), ($R_{31} - R_{13}$), and ($R_{23} - R_{32}$)
and set each expression equal to 0, then we produce  
\[    \nabla\times \mathbf{E} +\frac{1}{c} \frac {\partial }{\partial t} \, \mathbf{B}
 = 0.
   \]

\section{Other Equations of Electromagnetism}

Although the hypothesis relating three of Maxwell's equations to symmetry properties 
of the Ricci tensor served as the central guide to this calculation, other 
equations were reviewed.  
The expression $\mathbf{\nabla\cdot B}$ was not found in the Ricci tensor.  
A separate geometric 
source was found for this last member of Maxwell's equations.  
The continuity equation 
connects Coulomb's law with Ampere's law.  No  
geometric source for the continuity equation has been identified.  
The transformation properties of the connection were studied.  
With sufficient restrictions, the Lorentz transformation can be seen.

\subsection{$\mathbf{\nabla\cdot B = 0}$ }

The expression $\mathbf{\nabla\cdot B}$ was 
found in the Torsion tensor.
The Torsion is calculated directly from the connection as 
$T^{\,i}{}_{\!jk} = \Gamma^{i}{}_{\!jk} -\Gamma^{i}{}_{\!kj}$.
The Torsion tensor, shown in Table 3, depends on the magnetic field. 
If the Torsion were 
required to be zero, all connection components which depend 
on the magnetic field would be zero.

\begin{center}Table 3\\ Components of the Torsion $T$\\ showing only $j<k$\\[0.5ex]

\begin{tabular}{cccc|cccc|cccc}

 \textit{i} & \textit{j} & \textit{k} & $T^{\,i}{}_{\!jk}$ & \textit{i} & \textit{j} & 
    \textit{k} & $T^{\,i}{}_{\!jk}$ 
    & \textit{i} & \textit{j} & \textit{k} & $T^{\,i}{}_{\!jk}$ \\[0.5ex]
\hline
 & & & &   & & & &   & & & \\
2 & 0 & 3 & $ qB_x/mc^2 $ & 3 & 0 & 1 & $ qB_y/mc^2 $ & 1 & 0 & 2 & $ qB_z/mc^2 $ \\[0.2ex]
3 & 0 & 2 & $ qB_x/mc^2 $ & 1 & 0 & 3 & $ qB_y/mc^2 $ & 2 & 0 & 1 & $ qB_z/mc^2 $ \\[0.2ex]
0 & 2 & 3 & $ qB_x/mc^2 $ & 0 & 1 & 3 & $ -qB_y/mc^2$ & 0 & 1 & 2 & $ qB_z/mc^2 $ \\[0.2ex]

\end{tabular}

\end{center}

Define the Kronecker $\epsilon$ function as
\[ \epsilon_{ijkl} = \left\{ \begin{array}
{r@{\quad:\quad}l}
    0 & \{ i,j,k,l \}\quad \mbox{is not a permutation of} \quad  \{ 0,1,2,3 \}  
\\ +1 & \{ i,j,k,l \} \quad\mbox{is an even permutation of}\quad \{ 0,1,2,3 \} 
\\ -1 & \{ i,j,k,l \} \quad\mbox{is an odd permutation of} \quad \{ 0,1,2,3 \}  
\end{array} \right. 
\]
Then the Kronecker-signed sum of the components of the covariant 
derivative of the Torsion produces
\[
\sum_{ijkl} \epsilon_{ijkl}\, T^{\,i}{}_{\!jk,\,l}=  2  \, \frac{q}{mc^2} \, 
   \left( \frac {\partial }{\partial x} \,B_x 
+        \frac {\partial }{\partial y} \,B_y
 +       \frac {\partial }{\partial z} \, B_z    \right).
\]
Setting the sum to 0 yields the fourth of Maxwell's equations, $\nabla\mathbf{\cdot B}=0$ .

\subsection{Continuity Equation }

The continuity equation is $\partial \rho / \partial t + \nabla \mathbf{\cdot J} =0$.  
With the expressions for $\rho$ and $\mathbf{J}$ from 
Coulomb's law and Ampere's law, 
the equation becomes $\partial u / \partial t + \nabla \mathbf{\cdot S} =0$.  
Poynting's Theorem can be formulated as 
$\partial u / \partial t + \nabla \mathbf{\cdot S} = -\mathbf{J \cdot E}$,  
but $\mathbf{J \cdot E} \propto \mathbf{(E \times B) \cdot E} =0$ in this model. 
So $\partial u / \partial t + \nabla \mathbf{\cdot S} =0$ 
and the continuity equation is satisfied.

\subsection{Transformation properties of the connection: the Lorentz Transformation }

The Lorentz coordinate transformation for motion parallel to the $z$-axis is given by 
\[ x' = x\, , \qquad y'= y\, , \qquad z' =\gamma z - \gamma v t\, , \qquad t' = \gamma t  - \gamma v z /c^2, 
\]
where $\gamma = {(1-v^2/c^2)}^{\mbox{\bfseries -}1/2}$.  The  formula for the new 
connection components under such a coordinate transformation is complicated by   
a second-derivative 
term.  If we approximate the Lorentz transformation as a linear transformation 
with $\gamma $ and $v$ constant, then the 
second-derivative term goes away, and the connection transforms as a tensor.  
Even with this simplification, the expected transformation of 
the electric and magnetic fields is not observed for every 
connection component.

Looking at the classical Lorentz force converted to a geodesic format in 
equation \ref{lGeod}, there is one component based on $E_x$ and 
potentially four components based on $B_x$.  
We call components of this type ``observable'' components and   
conjecture that under a coordinate transformation the four ``observable'' 
$B$-components would not be seen individually, but rather 
as an average value.  For example, the transformed $E_x$ would be seen 
in the transformed individual component $-\Gamma^{1}{}_{\!00}$, 
while the transformed $B_x$ would be seen in the component average 
($\Gamma^{3}{}_{\!02} + \Gamma^{3}{}_{\!20} -(\Gamma^{2}{}_{\!30} + \Gamma^{2}{}_{\!03})$)/2.
This approach was found to be successful.

The effect of a transformation along the $z$-axis was calculated 
for the $x$-, $y$-, and $z$-field components.
Transformations for motion in the $x$ and $y$ directions give equivalent results.  
Terms of order $v^2/c^2$ are ignored  
in order to generate 
the expected transformed components. 
Note that within this approximation $\gamma\approx 1$;  excess powers of $\gamma$ were 
eliminated.  The result of the 
Lorentz coordinate transformation along the $z$-axis on the 
connection expressions is given in Table 4,
where the six ``observable'' component averages are displayed in the center column. 
The result agrees with the transformed electromagnetic 
fields displayed in Jackson~(1962) p.~380.

\begin{center}Table 4\\ Transformation of $\Gamma$ Components under a\\ Lorentz Coordinate Transformation 
along the $z$-axis\\[1.5ex]

\begin{tabular}{c |c|c}

 Initial Field & Average ``Observable'' Component & Transformed Field   \\[0.5ex]
\hline
{}& {} & {}\\
 $B_x  $&  $(\Gamma^{3}{}_{\!02} + \Gamma^{3}{}_{\!20} -(\Gamma^{2}{}_{\!30} + \Gamma^{2}{}_{\!03}))/2  $
&$\gamma B_x +\gamma \frac{v}{c} E_y  $\\[1.5ex]
 $ B_y  $&   $(\Gamma^{1}{}_{\!03} + \Gamma^{1}{}_{\!30} -(\Gamma^{3}{}_{\!10} + \Gamma^{3}{}_{\!01}))/2  $
&$\gamma B_y -\gamma \frac{v}{c} E_x  $   \\[1.5ex]
 $B_z  $&   $(\Gamma^{2}{}_{\!01} + \Gamma^{2}{}_{\!10} -(\Gamma^{1}{}_{\!20} + \Gamma^{1}{}_{\!02}))/2  $
&$B_z   $   \\[1.5ex]
 $ E_x  $&   $-\Gamma^{1}{}_{\!00}  $&$ \gamma E_x -\gamma \frac{v}{c} B_y $ \\[1.5ex]
 $E_y   $&   $-\Gamma^{2}{}_{\!00}  $&$\gamma E_y +\gamma \frac{v}{c} B_x  $  \\[1.5ex]
 $E_z  $&   $-\Gamma^{3}{}_{\!00}  $&$E_z   $ \\[1.5ex]

\end{tabular}

\end{center}

\section{The first Chern class}

The curvature tensor with components defined 
in equation \ref{defCurv} can be interpreted as a curvature 2-form 
matrix with elements
\[ \Theta^{\,i}{}_{\!j} = \sum_{k,l>k} R^{\,i}{}_{\!jkl}\: dx_{k}\wedge dx_{l}.
\] 
Neglecting a constant multiplier, the trace of the curvature 
2-form   $ \Theta^{i}{}_{\!i}$  is the first Chern class.  
Further ignoring the common multiplier $q/m c^2 $, the first Chern class is calculated to be 
\[ \frac{\partial E_x }{\partial t } \, dx \wedge dt 
+  \frac{\partial E_y }{\partial t } \, dy \wedge dt 
+  \frac{\partial E_z }{\partial t } \, dz \wedge dt
\]
\begin{equation} \label{trueChern}
-  \left(  \frac{\partial E_z }{\partial y }  - \frac{\partial E_y }{\partial z } \right) \, dy \wedge dz 
+  \left(  \frac{\partial E_x }{\partial z }  - \frac{\partial E_z }{\partial x } \right) \, dx \wedge dz 
-  \left(  \frac{\partial E_y }{\partial x }  - \frac{\partial E_x }{\partial y } \right) \, dx \wedge dy.      
\end{equation}
With Faraday's law, $\mathbf{\nabla \times E = - \frac{1}{c} \partial B} /\partial t$, 
and the definitions of the electromagnetic 
vector potential $\mathbf{A }$ and scalar potential $\Phi$, 
this 2-form can be shown to be 
the coboundary of a 1-form.  
Therefore, the 2-form in equation \ref{trueChern} is a 
member of the zero cohomology class and does not offer 
any interesting information about the universe.

\section{Lorentz Force }

The observed motion of a test particle follows the geodesic equation.
This equation in terms of the connection components is 
\[ \frac{d^{2}x_{i}}{ds^{2}} +\sum_{j,k} \Gamma^{i}{}_{\!jk} \frac{dx_{j}}{ds} \frac{dx_{k}}{ds} =0
\]
where $s$ is the parameter of the path.
Taking the real parts of the connection components from Table 2, 
the three ``space-like'' equations  are 
\[ m\, \frac{ d^2 x }{ d {s}^2} = E_x q { \frac{dt}{ds} }^{\mathbf{2}} +B_z \frac{q}{c} \frac{dy}{ds} \frac{dt}{ds} - B_y \frac{q}{c} \frac{dz}{ds} \frac{dt}{ds} -E_x  \frac{q}{c^2}  {\frac{dy}{ds}}^{\mathbf{2}}  -E_x \frac{q}{c^2}  {\frac{dz}{ds}}^{\mathbf{2}}     
\]
\[ m\, \frac{ d^2 y }{ d {s}^2} = E_y q { \frac{dt}{ds} }^{\mathbf{2}} +B_x \frac{q}{c} \frac{dz}{ds} \frac{dt}{ds} - B_z \frac{q}{c} \frac{dx}{ds} \frac{dt}{ds} -E_y \frac{q}{c^2}   {\frac{dx}{ds}}^{\mathbf{2}}  -E_y \frac{q}{c^2}  {\frac{dz}{ds}}^{\mathbf{2}} 
\]
\[ m\, \frac{ d^2 z }{ d {s}^2} = E_z q { \frac{dt}{ds} }^{\mathbf{2}} +B_y \frac{q}{c} \frac{dx}{ds} \frac{dt}{ds} - B_x \frac{q}{c} \frac{dy}{ds} \frac{dt}{ds} -E_z \frac{q}{c^2}   {\frac{dx}{ds}}^{\mathbf{2}}  -E_z \frac{q}{c^2}  {\frac{dy}{ds}}^{\mathbf{2}} 
\]

Making the change to proper time and defining
\[  s=c\, \tau \; , \qquad \frac{ d^2 x_i }{ d {\tau}^2} = a_i \; , \qquad
 \frac{ d x_i }{ d \tau} = v_i   \; ,  \qquad
  \frac{d t }{d\tau} \approx 1\, .
\]
this  simplifies to well-known vector equations.
In a magnetic field, the equation has the classical form
\[  \mathbf{F} = \frac{q}{c} \mathbf{v}\times \mathbf{B} \, .
\]
In a parallel electric field, it is the expected 
\[ \mathbf{F_{\parallel}} = q \mathbf{E_{\parallel}}   \, .
\]
However, in a transverse electric field, the force is 
different from the classical form at relativistic 
velocities with the value
\[  \mathbf{F_{\perp}} = q \mathbf{E_{\perp}} \,  (1 - v^2/c^2)\, .
\]

A new extension of the Lorentz force is 
\[  \frac{d^2\, t}{d\tau^2 }= 2 \frac{q}{mc^2} \left( \mathbf{ E\, \cdot} 
    \frac{d\mathbf{x}}{d\tau} \right)  \,\frac{dt}{d\tau}\, 
\]
Time will accelerate or decelerate relative to the proper time of a charged 
particle moving in an electric field.  Unstable charged particles 
moving in opposite directions 
parallel to an electric field should exhibit different decay rates.

\section{References}

\noindent Fock, V. 1927 On the Invariant Form of the Wave and Motion
Equations for a Charged Point-Mass[1]
\textit{Zeit. f. Physik} {\bfseries{39}} 226

\noindent Jackson, J. D. 1962 \textit{Classical Electrodynamics} New York: Wiley

\noindent Kaluza, T. 1921 On the Unification Problem in Physics 
\textit{Sitzungsber. Preuss. Akad. Wiss. Berlin} 966

\noindent Klein, O. 1926 Quantum Theory and Five-Dimensional Relativity 
\textit{Zeit. f. Physik} {\bfseries{37}} 895

\noindent O'Raifeartaigh, L. 1997 \textit{The Dawning of Gauge Theory}
Princeton University Press

\end{document}